\documentclass[conference]{IEEEtran}
\usepackage[utf8]{inputenc}
\usepackage{graphicx}
\usepackage{amsmath}

\usepackage[usenames,dvips]{color}
\usepackage{multirow}
\usepackage{booktabs}
\usepackage{amsmath}
\usepackage{amsfonts}
\usepackage{multirow}
\usepackage{epsfig}
\usepackage{psfrag}
\usepackage{subfigure}
\usepackage{pifont}
\usepackage{cite}
\usepackage{url}
\newcounter{numberlistc}
\usepackage{algorithm}
\usepackage{algorithmic}
\usepackage{graphicx}
\graphicspath{ {./figs/} }

\usepackage[usenames]{color}
\usepackage{amssymb,amsmath}
\usepackage{amsthm}
\usepackage{subfigure}

\usepackage{amssymb} 
 \usepackage{amsmath}

\newcounter{itemlistc}

\usepackage{setspace}

\usepackage[font=small, labelfont=bf]{caption}

\begin{document}

\title{Real-Time Differential Epidemic Analysis and Prediction for
  COVID-19 Pandemic}

%for conference 
\author{\IEEEauthorblockN{Sheldon X.-D. Tan\IEEEauthorrefmark{1} and
    Liang Chen\IEEEauthorrefmark{1}}
  \IEEEauthorblockA{\IEEEauthorrefmark{1}Department of Electrical and Computer
    Engineering, University of California, Riverside,  92521}
  \IEEEauthorblockA{ Emails: stan@ece.ucr.edu and lianchen@ucr.edu}
  }
  %\thanks{Corresponding author:  Sheldon Tan at stan@ece.ucr.edu}
%}
% for journal
% \author{
%   \indent Sheldon X.-D. Tan and Liang Chen
% ~\thanks{
%  \indent Sheldon X.-D Tan and Liang Chen  are with the Department of
%  Electrical and Computer Engineering,  University of California,
%  Riverside, CA 92521 USA. Please contact Sheldon Tan at
%  stan@ece.ucr.edu for any comments and question.
% }
%}

\maketitle

\begin{abstract}
  In this paper, we propose a new real-time differential virus
  transmission model, which can give more accurate and robust
  short-term predictions of COVID-19 transmitted infectious disease
  with benefits of near-term trend projection.  Different from the
  existing Susceptible-Exposed-Infected-Removed (SEIR) based virus
  transmission models, which fits well for pandemic modeling with
  sufficient historical data, the new model, which is also SEIR based,
  uses short history data to find the trend of the changing disease
  dynamics for the infected, the dead and the recovered so that it can
  naturally accommodate the adaptive real-time changes of disease
  mitigation, business activity and social behavior of
  populations. Our work is inspired by the observation that {\it
    contagious disease transmission prediction is similar to weather
    prediction: only short term prediction is typically accurate due
    to many time-varying interplaying factors as well as social and
    behavior uncertainties involved}.  On the other hand, accurate
  short-term prediction such as one week or ten days can give local
  government and hospical decision makers sufficient lead time for
  healthcare resources and critical personnel planning to provide
  in-patient treatment. As the parameters of the improved SEIR models
  are trained by short history window data for accurate trend
  prediction, our differential epidemic model, essentially are
  window-based time-varying SEIR model. Since SEIR model still is a
  physics-based disease transmission model, its near-term (like one
  month) projection can still be very instrumental for policy makers
  to guide their decision for disease mitigation and business activity
  policy changes in a real-time. This is especially useful if the
  pandemic lasts more than one year with different phases across the
  world like 1918 flu pandemic.  Numerical results on the recent
  COVID-19 data from China, Italy and US, California and New York
  states have been analyzed. A dedicated website has been built to
  show the projections based on the latest
  data~\cite{covid19_projection}.
  % Based on the projection as of May 1, 2020 from the
  % proposed model, the time for peak medical resource usage for US as a
  % whole will be around  the end of May. The total cumulative infected cases will reach to
  % peak of 1.799 million people around July 26, 2020 and the estimated
  % total death cases will reach to 112K in the end.
\end{abstract}

\section{Introduction}
  
The novel coronavirus (COVID-19) epidemic is generating significant
social, economic, and health impacts and has highlighted the
importance of real-time analysis and prediction of emerging infectious
diseases and health care resource and personnel planning and
economical activity guideling. 

One of the well-known models that is reasonably predictive for
human-to-human transmission is the so-called
Susceptible-Infectious-Removed (SIR) model, which was published in its
first form around the 1920s~\cite{weiss2013Mat2}. The model later has
been extended to consider more complicated situations such as the
Susceptible-Exposed-Infected-Removed (SEIR)
model~\cite{AndersonMay:Book'1991}. Those models are very successful
to describe how the disease dynamics will change over time once more
sufficient data are available (from outbreak to finish). Recently
those models have been applied to study recent COVID-19 transmissions
in different countries with certain
successes~\cite{Nesteruk:Medrxiv'2020,
  Chen:arXiv'2020,Peng:Medrxiv'2020, Zhou:JEBM'2020,
  Maier:Science'2020}.  For example, it is observed that city-wide
lockdown can lower the transmission rate substantially from those
models.  On the other hand, the data-driven and curve-fitting methods
for the prediction of COVID-19 such as Gaussian function based fitting
method in IHME projection~\cite{IHME_projection}, exponential curving
fitting in\cite{Zhao:IJID'2020}, machine-learning based approaches
in~\cite{zeng:arXiv'2020, hu:arXiv'2020} can fit the data well.
However, those methods generally suffer the lack of physical insights
of transmission and will not work well when the data does fit their
models well (For instance, two waves of infections). For real time
prediction, large projection ranges have to be given, which renders
the projection less valuable for medical resources and mitigation
policy planning.

Furthermore, the worldwide public health crisis like 1918 pandemic,
which killed an estimated 50 million people worldwide, including an
estimated 675,000 people in the United States, are difficult to model,
even to project~\cite{Jordan:CDC1918flu'2019}.  H1N1 flu lasted from
April 1917 to April 1919 for two years with three major phases across
different parts of world. Cities like San Francisco even experienced
strong second waves of death increase when the social distance was
relaxed too early.  Such long-term multi-phase and multi-year
transmission dynamic is very difficult to be captured by existing
pandemic models.   As a result, real-time short-term prediction and
near-tern trend projection can give each city and county policy maker
and resource planners extremely valuable information to guide the
disease mitigation and business open/close decision in a real time.

On the other hand, traditional physics-based SIR/SEIR epidemic
modeling and its variants suffer several drawbacks especially for
real-time disease transmission predictions as it is static model in
which many key parameters such as transmission rate and recovery rates
are typically fixed values from fitting.  However, for countries like
US, the local government interventions and mitigation policies such as
active surveillance, contact tracing, quarantine, massive testing,
school and business closure, shelter in place, social distancing
etc. keep changing to respond the local epidemic situations. Also for
each state in US has different time lines for implementing different
prevention and mitigation measures, which make the static based
prediction even more difficult to predict.  Further the social
behaviors of population, such as bearing face coverage masks are not
consistent through different regions and cities as the pandemic
progresses, which will affect the transmission rate as well.  The
availability of healthcare resources of each city and county, which
are also changing factors, which may also affect recovery rate.  As a
result, existing static SEIR models do not fit well for real-time
disease prediction as the key parameters such as transmission rate,
recovery rate, which is also closely related to the effective
reproduction rate $R0$, are time-varying parameters. On the other
hand, one important observation is that {\it contagious disease
  transmission prediction is similar to weather prediction: only short
  term prediction is accurate due to many time-varying interplaying
  factors as well as social and behavior uncertainties involved}.  The
widely watched IHME (Institute for Health Metrics and Evaluation)
model for real-time COVID-19 epidemic
prediction~\cite{IHME_projection} is purely based on mathematic curve
fitting techniques, which lacks the theoretical foundation of epidemic
transmission found in the SIR models and thus its prediction accuracy
is highly debatable.

To mitigate this problem, time-varying SEIR models have been proposed
in the past.  Dureau {\it et al} tried to model the time-varying
affects of the SEIR models by considering partial and noisy data. They
introduced stochastic processes into the SEIR models and solved the
resulting stochastic SEIR partial differential equations using Markov
Chain Monte Carlo methods in which the transmission is modeled as
random walks, which are very expensive to
compute~\cite{Dureau:Biostat'2013}. This approach was applied to study
the transmission within and outside Wuhan for January to February
2020~\cite{Kucharski:Lancet'2020db}.  Recently Chen {\it et al}
proposed time-dependent SIR model to model the COVID-19
outbreaks in China~\cite{Chen2020ATS}, In this mode, model parameters
are computed in a daily basis. As a result, it lacks the good
predictability as all the key parameters have been predicted a prior
first, and some ad-hoc methods was introduced to predict the
parameters in a near future.

In this work, we propose a new real-time differential virus
transmission modeling method, which can give more accurate and robust
short-term predictions of COVID-19 transmitted infectious disease
while still maintain the near-term projection benefit.  The new model
is based on enhanced Susceptible-Exposed-Infected-Removed (SEIR) virus
transmission model. But it tries to obtain the differential view of
pandemic dynamics in a short history window to analyze the short trend
of the transmission so that it can be more accurate over different
periods of time to accommodate the adaptive changes of disease
mitigation, business activity and social behavior of populations. As
the parameters of the improved SEIR models are trained by short
history window data for accurate trend prediction, our differential
epidemic model, essentially are window-based time-varying SEIR
model. Since SEIR model still is a physics-based disease transmission
model, its near-term (like one month) projection can still be very
instrumental for policy makers to guide their decision for disease
mitigation and business activity policy changes in a real-time.
Numerical results on the recent COVID-19 data from China, Italy and
US, California and New York states are analyzed. A dedicated website
has been built to show the projections based on the latest
data~\cite{covid19_projection}.

% Based on the projection as of May 1, 2020, the US
%  actively infected cases will reach the peak
% around the end of May and total cumulative infected cases will
% reach to peak around July 26 with 1.799 million people and the
% estimated total death cases will reach to 112K in the end.

\begin{figure}[h]
\centering
\includegraphics[width=0.81\columnwidth]{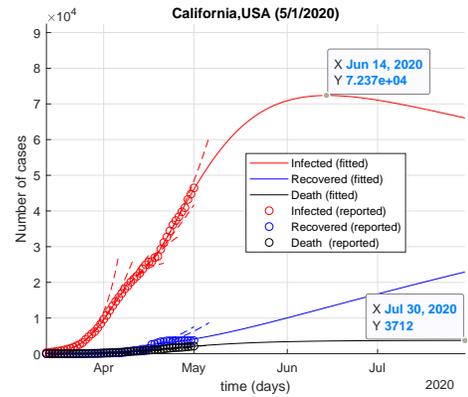}
\caption{Modeling and predictions of the proposed differential SEIRDP
  model for California from early Match to later July 2020 }
\label{fig:California_SEIRDP}
\end{figure}

Fig.~\ref{fig:California_SEIRDP} illustrates the modeling and
prediction results from the proposed differential SEIR model for the
COVID-19 disease for California from early March to later July
2020. As we can see, at the beginning, the infected grows very fast,
the projected growth also reflects such fast changing rates. But as
social distancing and stay at home policies were introduced across
cities and counties in California in the later March, the growth rate
went down, the projected growth at different time point also reflects
such trends. Based on the projection of our model at May 1, 2020, the
infected case will reach to the peak about 72.37K around Jun 14 in
California, which indicates the peak medical resources needed. In
contrast, the well-watched IHME's prediction~\cite{IHME_projection}
predicts that the peak medical resource needed is around April 17.  We
will show more significant differences between our models and IHME's
prediction ~\cite{IHME_projection} for New York state later.

 \section{The enhanced SEIR base model}
\begin{figure}[h]
\centering
\includegraphics[width=0.81\columnwidth]{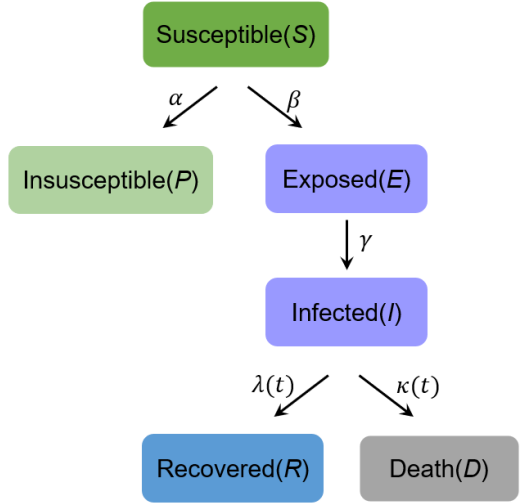}
\caption{The proposed enhanced SEIR  base model}
\label{fig:enhanced_seir_model}
\end{figure}

 In this section, we first present the extended SIR model, which is a
 extension of classical SEIR
 model~\cite{Tang:JCM'2020,Tang:IDM'2020,Labadin:Medrxiv'2020,Shen:Medrxiv'2020,Clifford:Medrxiv'2020,Xiong:Medrxiv'2020,Li:Medrxiv'2020}
 as the base model for the proposed differential modeling shown later.
 The proposed base model, called {\it SEIRDP model}, is similar to the
 recently proposed generalized SEIR model for studying the COVID-19
 disease in Wuhan and China~\cite{Peng:Medrxiv'2020}.  We removed the
 quarantine compartment in the proposed SEIRDP model as there is no
 quarantine data for most of countries outside China. The resulting
 SEIRDP (Susceptible, Exposed, Infected, Recovered, Death,
 Insusceptible (P)) model is shown in
 Fig.~\ref{fig:enhanced_seir_model}.  In this base model, we have six
 states, i.e.  $\{S(t), P(t), E(t), I(t), R(t), D(t)\}$, which
 indicate the number of the susceptible cases, insusceptible cases,
 exposed cases (infected but not yet be infectious, in a latent
 period), infectious cases (confirmed with infectious capacity),
 recovered and immune cases and closed cases (or death). Then the
 total number of population in a certain region or county is
 $N = S+P +E+I +R+D$. The coefficients
 $\{ \alpha, \beta, \gamma^{-1}, \lambda^{-1}, \kappa \}$ represent
 the protection rate, infection rate, average latent time, cure rate,
 and mortality rate, respectively. The introduction of insusceptible
 compartment represents gradual changing (or growing) population,
 which will not be infected due to some strong disease mitigation
 measures such as enforced shelter in place, strict city lockdown in
 China and European countries. The basic reproduction number, $R0$,
 represents the number of secondary infections from a primary infected
 individual in a fully susceptible population, which can be computed
 by~\cite{Peng:Medrxiv'2020}
\begin{equation}
  R0 = \beta \lambda^{-1} (1 - \alpha)^n
  \label{eq:r0_old}
\end{equation}
where $n$ is the number of days. 
 We remark that If we force $\alpha =0$, the proposed SEIRDP model
 essentially becomes the classical SEIR model (if we put recovered and
 death cases into one recovery compartment).
 
\section{The proposed differential SEIR model}
Once we have the base SEIR model, then we can present our differential
SEIR model. In this model, basically the key parameters will become
time dependent. 

\begin{figure}[h]
\centering
\includegraphics[width=0.81\columnwidth]{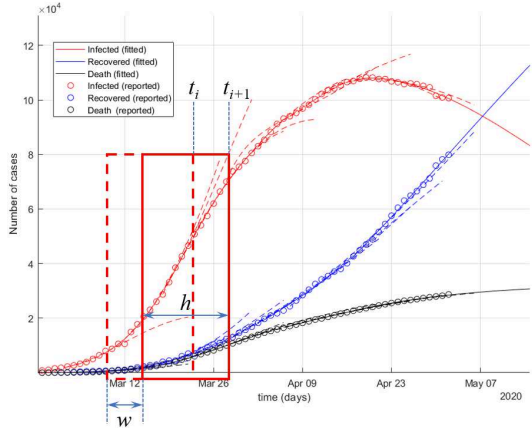}
\caption{Fitting in a moved time window}
\label{fig:window}
\end{figure}

In our work, we introduce a time window concept, which represents a
short period of time in a few days, as shown in Fig.~\ref{fig:window}.
In this figure, the red box present a history length of the data ($h$)
we use for training the model. The red box will move one window size
($w$) as one step forward. For instance five days can be a
good time window. In each time window, all the key parameters in the
base SEIR models are constant. This reflects the fact that the
transmission and recovery conditions for population does not change in
a short window time so that we can use a traditional static SEIR model
based on the history of data around this window time and we can also
predict a short future with sufficient accurate assuming such disease
dynamic does not change dramatically, which is also reasonable.  Based
on this observation, we can present the resulting ordinary differential
equations for the proposed differential SEIR model as follows:

\begin{equation}
    \begin{split}
        &\frac{d S(t)}{d t}=-\beta(t_i) \frac{S(t) I(t)}{N}-\alpha(t_i) S(t)\\
        &\frac{d E(t)}{d t}=\beta(t_i) \frac{S(t)
          I(t)}{N}-\gamma(t_i)E(t)\\
        &\frac{d I(t)}{d t}=\gamma(t_i)E(t)-\lambda(t_i,t)I(t)-\kappa(t_i,t)I(t)\\
        &\frac{d R(t)}{d t}=\lambda(t_i,t) I(t)\\
        &\frac{d D(t)}{d t}=\kappa(t_i,t) I(t)\\
        &\frac{d P(t)}{d t}=\alpha(t_i)  S(t)
      \end{split}
      \label{eq:enh_seir_pde}
    \end{equation}
    where time $t$ belongs to the $i$th time window.  If we use $t_i$
    to indicate the last day in the window $i$, then we have
    $t \in [t_i -h + 1, t_i]$, where $h$ is the window size. 
    The window can be moved forward with stride $w$ days. 
    Therefore, we have the relationship $t_{i+1}=t_i+w$. The cure
    rate $\lambda(t_i, t)$ and mortality rate $\kappa(t_i, t)$ are
    time-dependent, which again are dependent on two parameters. They
    again are time-window dependent (explained below):
\begin{equation*}
    \begin{split}
        &\lambda(t_i, t) = \lambda_0(t_i) [1-\exp(-\lambda_1(t_i)  t)]\\
        &\kappa(t_t, t)) = \kappa_0(t_i)  \exp(-\kappa_1(t_i)  t)\\
    \end{split}
  \end{equation*}
  Those two equations basically say that cute rate or recovery rate
  $\lambda(t)$ will goes to constant value $\lambda_0$ exponentially
  over a time and the mortality rate $\kappa(t)$ goes down
  exponentially over time~\cite{cheynet:2020}.

  We notice that all the key parameters for the enhanced SEIR model
  are time-varying instead of fixed values.  But they are so-called
  {\it time window dependent} as they do not change every day compared
  to existing time-varying based SEIR models. Specifically, we select
  $w=5$ days as a window (the window size is a hyper parameter and can
  be optimized for different regions and countries). We perform each
  prediction over each non-overlapping window and then slide one
  window forward for next prediction and so on. As a result, within
  the $i$th window, the seven parameters, $\{ \alpha(t_i)$,
  $ \beta(t_i)$, $\gamma^{-1}(t_i)$, $\lambda_0^{-1}(t_i)$,
  $\lambda_1^{-1}(t_i)$, $\kappa_0(t_i)$,$\kappa_1(t_i)\}$ are kept
  constant. The parameters will be found by a regression process shown
  next section.  The stride can be $h= 10$ days or $h=15$ days,
  which is also another hyper parameter for our model.  Since all the
  parameters are time window dependent, the reproduction number
  becomes time dependent, which is also called {\it effective
    reproduction number}, $R_t$ at time $t$ in $i$th time window.
  $R_t$ can be computed as follows:
  \begin{equation}
  R_t (t_i) = \beta(t_i) \lambda^{-1} (t_i)\prod_{j=1}^{n}(1 - \alpha_j)
  \label{eq:r0_new}
\end{equation}

We note that such window-based SEIR model obtained at $i$th window
also depends on the historical data from the previous windows.  The
reason is that our differential model can be viewed as performing the
differential operations on a dynamic systems at a specific time frame
or window, not on a static function. As a result, the impacts from the
historical data of the previous windows will be represented as the
initial conditions for solving the PDE of \eqref{eq:enh_seir_pde} and
obtaining the resulting parameters by fitting the resulting
discretized PDE with the data in this window.

\section{Window-dependent model parameter estimation for
  differential SEIR models}
For the proposed differential SEIR model, for each differential
window, we will calculate the seven $\{ \alpha(t_i)$, $ \beta(t_i)$,
$\gamma^{-1}(t_i)$, $\lambda_0^{-1}(t_i)$, $\lambda_1^{-1}(t_i)$,
$\kappa_0(t_i)$,$\kappa_1(t_i)\}$ in the $i$th time window indicated
by $t_i$, which indicates the last day in the window $i$.
As a result, we can rewrite the partial differential equation
\eqref{eq:enh_seir_pde} into the following initial value PDE in matrix form with the
initial condition $\mathbf{Y}_{0}$, which indicate the the impacts from the
historical data before $i$th window:
\begin{equation}
\begin{split}
  & \mbox{PDE}: \frac{d \mathbf{Y}}{d t} = \mathbf{A}\cdot\mathbf{Y}+\mathbf{F} \\
  & \mbox{IC}: \mathbf{Y}_{0}
\end{split}
\end{equation}
where {\it IC} is initial condition. 
\begin{equation*}
    \begin{split}
        &\mathbf{Y}=
        \begin{bmatrix}
        S(t) & E(t) & I(t) & R(t) & D(t) & P(t)\\
        \end{bmatrix}^T\\
        &\mathbf{A}=
        \begin{bmatrix}
        -\alpha(t_i) & 0 & 0 & 0 & 0&0\\
        0 & -\gamma(t_i) & 0 & 0 & 0&0\\
        0 & \gamma(t_i) & -\lambda(t_i,t)-\kappa(t_i,t) & 0 & 0&0\\
        0 & 0 & \lambda(t_i,t) & 0 & 0&0\\
        0 & 0 & \kappa(t_i,t) & 0 & 0&0\\
        \alpha(t_i) & 0 & 0 & 0 & 0&0\\
        \end{bmatrix}\\
        &\mathbf{F}= S(t)\cdot I(t) \cdot
        \begin{bmatrix}
        -\frac{\beta(t_i)}{N} & \frac{\beta(t_i)}{N} & 0 & 0 & 0 & 0\\
      \end{bmatrix}^T\\
       & \mathbf{Y}_{0}= \mathbf{Y}(t=t_i-h+1)=\mathbf{Y}(t=t_{i-1}-h+w+1)  
    \end{split}
  \end{equation*}

  Then we perform the time domain discretization using simple Forward
  Euler or higher order explicit Runge-Kutta method and we end up of
  number of algebraic equations. By solving this resulting equation in
  the time domain, we can obtain the seven parameters over the given
  data of  $\{ I(t), R(t), D(t)\}$ in the past 10 or 15 days with the
  initial state conditions computed from the previous time window. In
  this paper, we use a nonlinear least-squares solver to estimate the
  parameters by the expression~\cite{cheynet:2020}
\begin{equation}
	\hat{\theta}=\arg \min _{\theta}:= ||\text{SEIRDP}(\theta, time)-
	\begin{bmatrix}
	I\\
	R\\
	D\\
	\end{bmatrix}
	||^2
\end{equation}
where $\theta = \{ \alpha(t_i)$, $ \beta(t_i)$, $\gamma^{-1}(t_i)$,
$\lambda_0^{-1}(t_i)$, $\lambda_1^{-1}(t_i)$,
$\kappa_0(t_i)$,$\kappa_1(t_i)\}$, $\hat{\theta}$ is the parameters of
estimated model, $time$ is the time from the real data, $\{I, R, D\}$
are the infected, recovered and death cases from the real data, and
SEIRDP$(\cdot)$ represents the SEIRDP model.

\section{Results and discussion}
\label{sec:results}

\subsection{Analysis for public data from  Hubei Province, China}

\begin{figure}[!ht]
  \centering
  \includegraphics[width=0.81\linewidth]{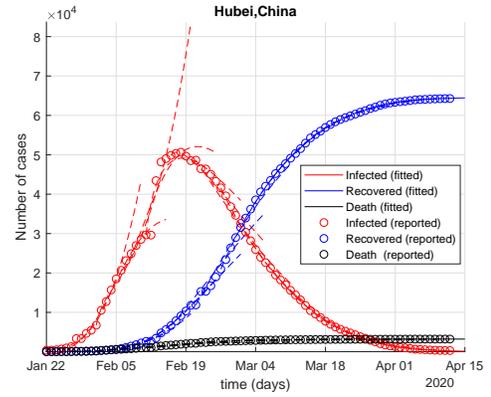}
  \caption{Predictions of the SEIRDP model for Hubei province, China
    from January to middle April, 2020}
\label{fig:SEIRDP_hubei_IRD}
\end{figure}

\begin{figure}[!ht]
  \centering
  \includegraphics[width=0.81\linewidth]{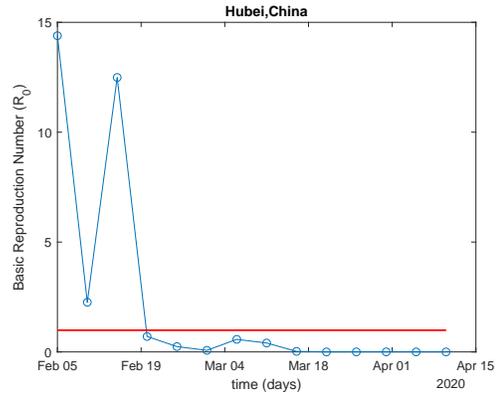}
  \caption{Effective reproduction number for Hubei province, China
    from January to middle April, 2020}
\label{fig:SEIRDP_hubei_BRN}
\end{figure}

\begin{figure}[h]
\centering
\includegraphics[width=0.81\columnwidth]{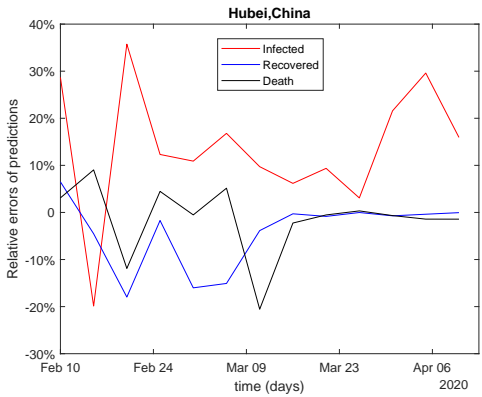}
\caption{The  relative mean errors of projection for Hubei province China from January to middle of
  April,  2020}
\label{fig:SEIRDP_hubei_error}
\end{figure}

We first show the modeling results for Hubei province of China from
early January to the middle of April, 2020.
Fig.~\ref{fig:SEIRDP_hubei_IRD} shows the time evolution of the
numbers of the currently infected, the recovered and the death cases
over this period. Note that many public websites show the cumulative
cases, which are different than the currently infected cases. The
currently infected cases equal the cumulative case minus the recovered
and dead cases.  The circles in the figure (in all the figures) are the
measured data and the solid line is the last project in early April
2020 as China's pandemic outbreak basically has run its course already
and we have all the historical data from early outbreak to finish.

Notice that on Feb. 12, 2020, COVID-19 testing criteria was relaxed in
Hubei Province. As a result, there is a huge jump of confirmed
infected cases, which can be viewed as an outlier for the date. As a
result, the projection based on the history before and around Feb. 12
is quite off the track of actual cases as shown in Fig.~\ref{fig:SEIRDP_hubei_IRD}. 
 But for the most of days over the two
month period, the projected and actual measured cases match very well.

Fig.~\ref{fig:SEIRDP_hubei_BRN} shows the effective reproduction
number, $R_t$, over the same period time. As we can see, $R_t$ becomes
less than 1 around Feb. 19, which is about 3 weeks after city lockdown
in Wuhan on Feb 23, 2020. This indeed shows the effects of strict city
lockdown in Hubei province to reduce the transmission effect of the
virus. Feb. 19 is also close to the time when the peak of infected
people was reached as shown in Fig.~\ref{fig:SEIRDP_hubei_IRD}.  As a
result, analysis and prediction of effective reproduction number can
give us more insights into how the pandemic dynamics will play out and
when the peak or turning points will happen in the real
time. Fig.~\ref{fig:SEIRDP_hubei_error} shows the 5-day predicted mean
errors in percentage  against the measured
data. The mean errors are computed for the average estimated errors in
each time window (every five days) between the 5-day projected
infected cases and measured cases. This is case for all the mean error
computations in the sequel.

% \begin{figure}[!ht]
%   \centering
%   \subfigure[]{\includegraphics[width=0.49\linewidth]{SEIRD_hubei_IRD}\label{fig:SEIRD_hubei_IRD}}
%   \subfigure[]{\includegraphics[width=0.49\linewidth]{SEIRD_hubei_BRN}\label{fig:SEIRD_hubei_BRN}}
%   \caption{(a)Predictions of the SEIRD model and (b)Effective reproduction number for Hubei province. 
%   \label{fig:Hubei_SEIRD}}
% \end{figure}

\subsection{Analysis and prediction for Italy}

\begin{figure}[!ht]
  \centering
  \includegraphics[width=0.81\linewidth]{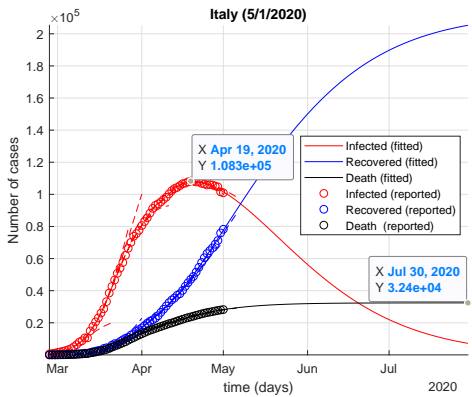}
  \caption{Predictions of the differential SEIRDP model for Italy from
    later February to middle April,  2020}
\label{fig:SEIRDP_Italy_IRD}
\end{figure}

\begin{figure}[!ht]
  \centering
  \includegraphics[width=0.81\linewidth]{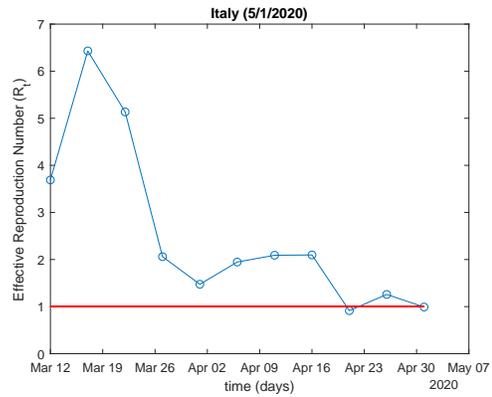}
  \caption{Effective reproduction number for Italy from early March
    to later July, 2020}
\label{fig:SEIRDP_Italy_BRN}
\end{figure}

\begin{figure}[h]
\centering
\includegraphics[width=0.81\columnwidth]{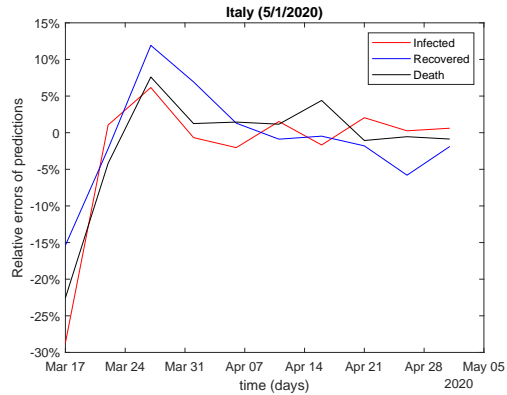}
\caption{The relative mean errors of projection for
  Italy from Februray to April,  2020}
\label{fig:SEIRDP_Italy_error}
\end{figure}

Fig.~\ref{fig:SEIRDP_Italy_IRD} shows COVID-19 disease modeling and
prediction for Italy from late February to middle of April 2020 as
Italy was mostly severely impacted by COVID-19 in Europe. As we can
see, our differential models match well for the historical data. The
prediction around early March is quite aggressive as the actual growth
rate for both infected and death are very high. This is also reflected
in the effective reproduction number, which is about 6-7, at those
days as shown in Fig.~\ref{fig:SEIRDP_Italy_BRN}.  As of April 13,
$R_t$ has reached to around 2 and is still going downward. Based on
our 10 day prediction, the currently infected cases will reach to the
peak around April 21 with about 107K cases.
Fig.~\ref{fig:SEIRDP_Italy_error} shows the 5-day projected mean
errors. As we can see, as time progresses, the projected error goes
down and is limited to about 10\% for the three types of cases
analyzed.

% \begin{figure}[!ht]
%   \centering
%   \subfigure[]{\includegraphics[width=0.49\linewidth]{SEIRD_Italy_IRD}\label{fig:SEIRD_Italy_IRD}}
%   \subfigure[]{\includegraphics[width=0.49\linewidth]{SEIRD_Italy_BRN}\label{fig:SEIRD_Italy_BRN}}
%   \caption{(a)Predictions of the SEIRD model and (b)Effective reproduction number for Italy. 
%   \label{fig:Italy_SEIRD}}
% \end{figure}

\subsection{Analysis and prediction for United State}
For US data, we show the results for US, New York state and California
state as New York is the epicenter of the COVID-19 in US. California
has the largest population in US. The two states also have dramatically
different transmission situations.

\begin{figure}[!ht]
  \centering
  \includegraphics[width=0.81\linewidth]{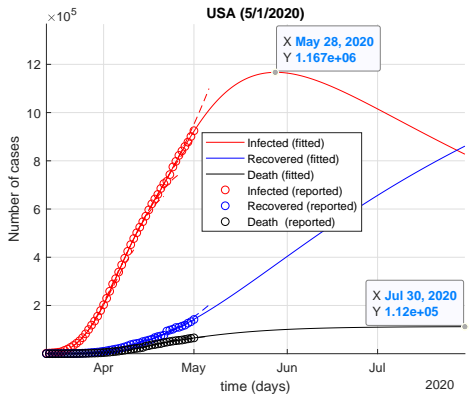}
  \caption{Predictions of the differential SEIRDP model for US from
    early March to middle July, 2020}
\label{fig:SEIRDP_US_IRD}
\end{figure}

Fig.~\ref{fig:SEIRDP_US_IRD} shows first modeling and projection for
US from late early March to middle  July 2020 for the currently
infected, recovery and death cases. The results from the models match
the given data very well.
% Based on the last differential SEIRDP model,
% we project that the infected cases in US will reach to the peak  at
% May 5th, 2020 with about 767K cases, which is also the peak time for
% medical resources. As comparison, the IHME's
% projection~\cite{IHME_projection} predicts the peak medical usage is
% around April 10, 2020.
% Based on the projection as of May 1, 2020 from the proposed model, the
% time for peak medical resource usage for US as a whole will be around
% the end of May. The total cumulative infected cases will reach to peak
% of 1.799 million people around July 26, 2020 and the estimated total
% death cases will reach to 112K in the end.

\begin{figure}[!ht]
  \centering
  \includegraphics[width=0.81\linewidth]{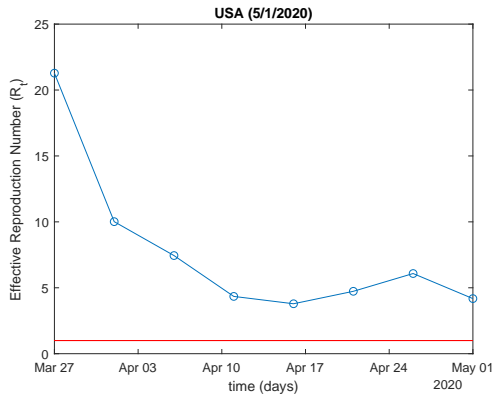}
  \caption{Effective reproduction number for US from early March to
    middle April, 2020}
\label{fig:SEIRDP_US_BRN}
\end{figure}

\begin{figure}[h]
\centering
\includegraphics[width=0.81\columnwidth]{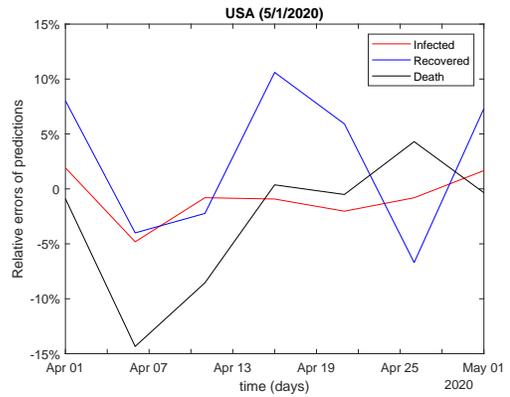}
\caption{The relative mean errors of projection  for
  US from early March to middle April, 2020}
\label{fig:SEIRDP_US_error}
\end{figure}

The effective reproduction numbers over the same time period and the
relative projection errors are shown in Fig.~\ref{fig:SEIRDP_US_BRN}
and Fig.~\ref{fig:SEIRDP_US_error} respectively.  As we can see, the
effective reproduction number seems higher than basic reproduction
number, $R0$, which was estimated about 2-7 for
COVID-19~\cite{Liu:JTM'2020}. This may reflect the significant portion of
unconfirmed and asymptotic population at the beginning of this
outbreak in US (this is the cases for all the other countries). But
as time progresses, $R_t$ tends to level down to a more reasonable
range as more confirmed cases have been reported.

\begin{figure}[h]
\centering
\includegraphics[width=0.81\columnwidth]{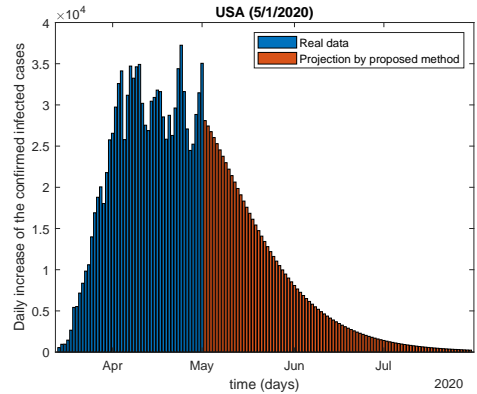}
\caption{The daily increase of infected cases  measured and projected for US from
  early Match to middle July, 2020}
\label{fig:SEIRDP_US_NewCase}
\end{figure}

\begin{figure}[h]
\centering
\includegraphics[width=0.81\columnwidth]{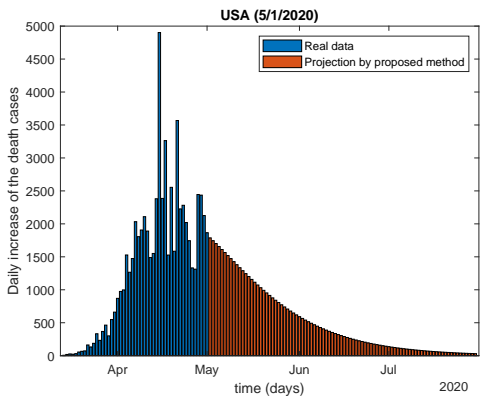}
\caption{The daily increase of death cases measured and  projected for US from early
  Match to middle July, 2020}
\label{fig:SEIRDP_US_NewDeath}
\end{figure}

Fig.~\ref{fig:SEIRDP_US_NewCase} and Fig.~\ref{fig:SEIRDP_US_NewDeath}
show that daily newly confirmed infected cases and daily death cases
for US over the same period respectively. The blue lines are the
actual data and the brown lines are projected data from April 13 to
middle of July 2020.
% As we can see that the daily infected cases
% seem reaching the peak around April 10 with about 35K per day and
% then start to de-escalate. From Fig.~\ref{fig:SEIRDP_US_NewDeath}, we
% can see that the daily death reached the peak around April 10 and then
% go downward gradually.  If this trend is held, the daily death will
% continue well beyond July before it stops and the estimated total
% death cases will reach about 90K in the end.
We want to express that
such projection is subject to change due to many changing factors such
as mitigation policies and people behaviors in the near term.

% We also report the projected numbers from the
% static base SEIRPD model to show the differences between the two
% models. As we can see the discrepancies are quite significance. The
% proposed differential models match better with the measured data
% around time for the real-time prediction (April 12, 2020).

\begin{figure}[h]
\centering
\includegraphics[width=0.81\columnwidth]{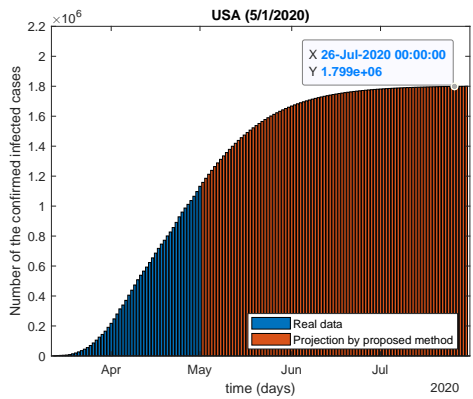}
\caption{The number of the confirmed cases measured and projected for US from early
  Match to middle July, 2020}
\label{fig:SEIRDP_US_Confirmed}
\end{figure}

Fig.~\ref{fig:SEIRDP_US_Confirmed} show the existing daily confirmed
infected 
cases for US over March to middle July. We project it will reach total
number probably will reach to about 1.1 millions  in June 12.

\subsection{Analysis and prediction for California state}
\begin{figure}[!ht]
  \centering
   \includegraphics[width=0.81\linewidth]{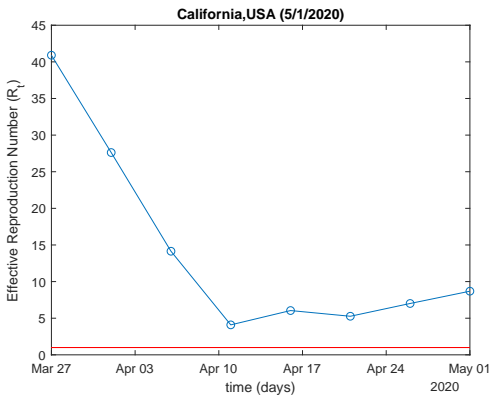}
   \caption{The  estimated effective reproduction number for California
     from early Match to middle April, 2020}
\label{fig:SEIRDP_California_BRN}
\end{figure}

\begin{figure}[!ht]
  \centering
    \includegraphics[width=0.81\linewidth]{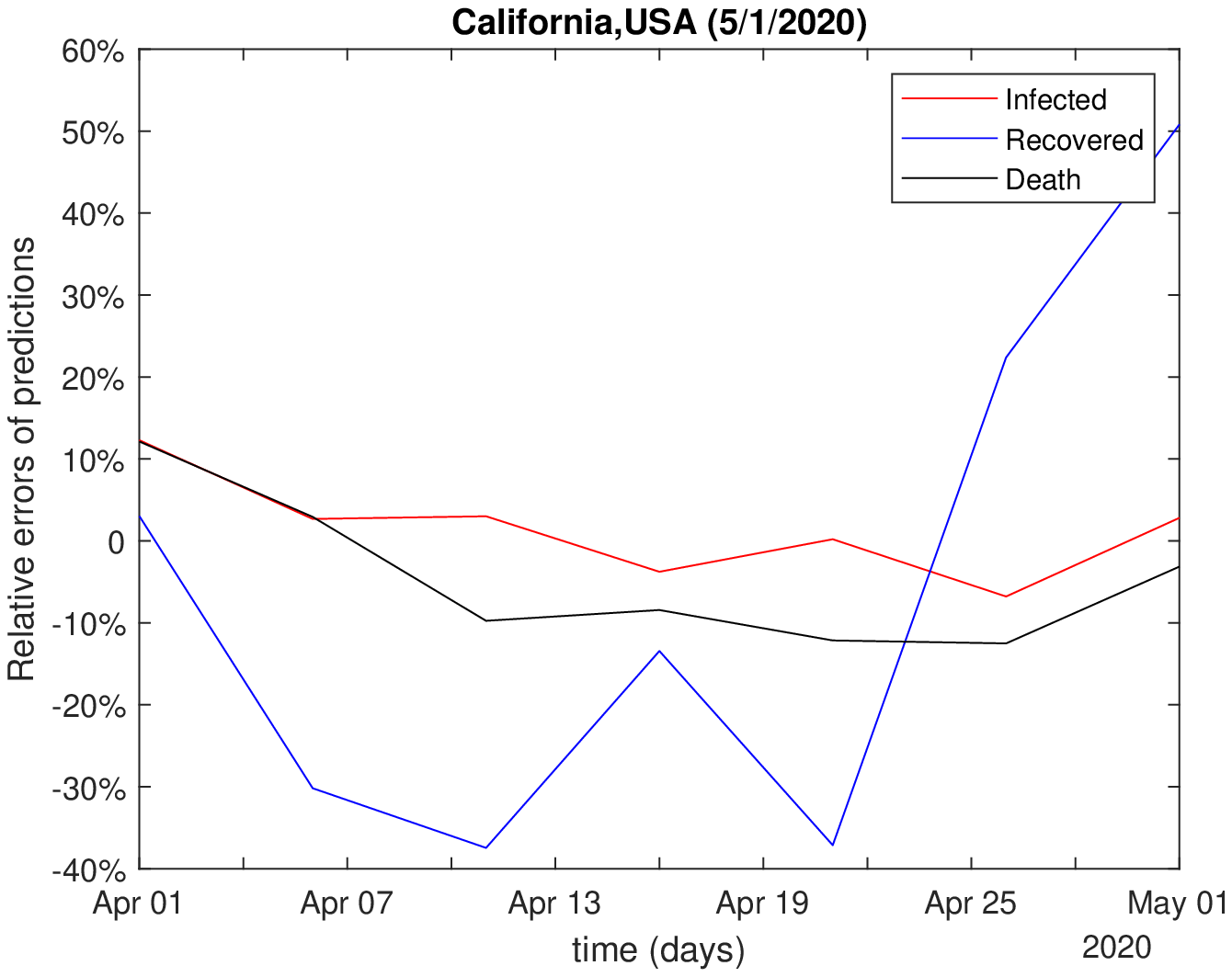}
    \caption{The relative mean errors of
      projection for California from early Match
      to middle April, 2020 }
\label{fig:SEIRDP_California_error}
\end{figure}

For California, the modeling and projection results from the same
period of US are shown in Fig.~\ref{fig:SEIRDP_US_IRD},
Fig.~\ref{fig:SEIRDP_California_BRN} and
Fig.~\ref{fig:SEIRDP_California_error}.

% As we can see from
% Fig.~\ref{fig:SEIRDP_US_IRD}, the currently infected cases will reach
% the peak (about 26.4K) around April 26. In contrast, IHME's
% projection~\cite{IHME_projection} predicts that the time for peak
% medical resource usage is around April 17.

The effective reproduction
number is around 5 as of April 13, which is close to the observed
basic reproduction number for COVID-19, estimated about 2-7
for COVID-19~\cite{Liu:JTM'2020}.

\begin{figure}[h]
\centering
\includegraphics[width=0.81\columnwidth]{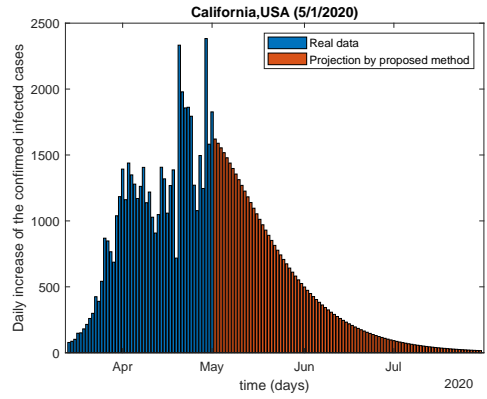}
\caption{The daily increase of infected cases measured and  projected for California from
  early Match to middle July, 2020}
\label{fig:SEIRDP_California_NewCase}
\end{figure}

\begin{figure}[h]
\centering
\includegraphics[width=0.81\columnwidth]{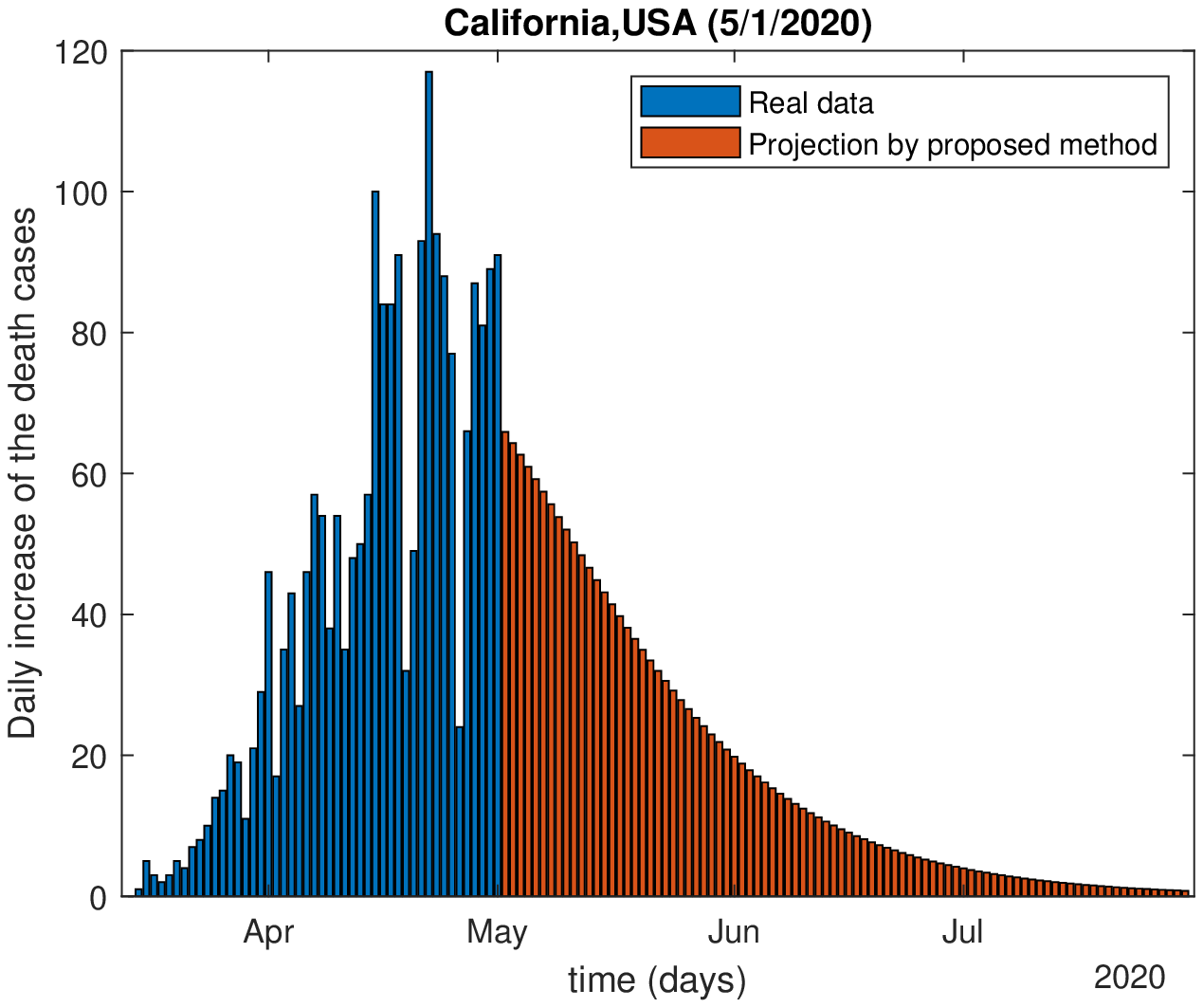}
\caption{The daily increase of death cases measured and projected for California from early
  Match to middle July, 2020}
\label{fig:SEIRDP_California_NewDeath}
\end{figure}

\begin{figure}[h]
\centering
\includegraphics[width=0.81\columnwidth]{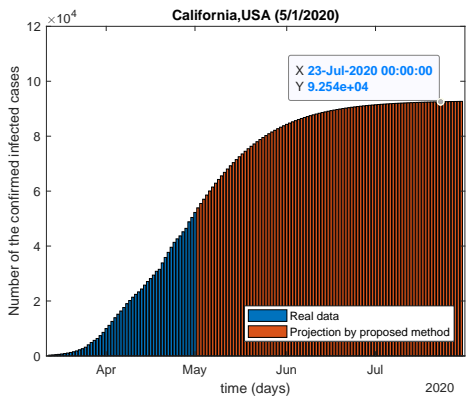}
\caption{The number of the confirmed cases  measured and projected for California from early
  Match to middle July, 2020}
\label{fig:SEIRDP_California_Confirmed}
\end{figure}

Fig.~\ref{fig:SEIRDP_California_NewCase},
Fig.~\ref{fig:SEIRDP_California_NewDeath} and
\ref{fig:SEIRDP_California_Confirmed} show the daily confirmed
infected cases, the death cases and the total accumulative infected
cases for California over mentioned period respectively.
% California
% also reached the peak on April 8 (about 14K) for daily infected and on
% April 10 for death cases (about 54).  But the total infected total is
% projected to reach 32.7K at May 23, and the estimated total death
% cases will reach to about 1585 in the end.

% \begin{figure}[!ht]
%   \centering
%   \subfigure[]{\includegraphics[width=0.49\linewidth]{SEIRD_US_IRD}\label{fig:SEIRD_US_IRD}}
%   \subfigure[]{\includegraphics[width=0.49\linewidth]{SEIRD_US_BRN}\label{fig:SEIRD_US_BRN}}
%   \caption{(a)Predictions of the SEIRD model and (b)Effective reproduction number for US. 
%   \label{fig:US_SEIRD}}
% \end{figure}

% \begin{figure}[!ht]
%   \centering
%   \subfigure[]{\includegraphics[width=0.49\linewidth]{SEIRDP_California_IRD}\label{fig:SEIRDP_California_IRD}}
%   \subfigure[]{\includegraphics[width=0.49\linewidth]{SEIRDP_California_BRN}\label{fig:SEIRDP_California_BRN}}
%   \caption{(a)Predictions of the SEIRDP model and (b)Effective reproduction number for California. 
%   \label{fig:California_SEIRDP}}
% \end{figure}

% \begin{figure}[!ht]
%   \centering
%   \subfigure[]{\includegraphics[width=0.49\linewidth]{SEIRD_California_IRD}\label{fig:SEIRD_California_IRD}}
%   \subfigure[]{\includegraphics[width=0.49\linewidth]{SEIRD_California_BRN}\label{fig:SEIRD_California_BRN}}
%   \caption{(a)Predictions of the SEIRD model and (b)Effective reproduction number for California. 
%   \label{fig:California_SEIRD}}
% \end{figure}

\subsection{Analysis and prediction for New York state}
\begin{figure}[!ht]
  \centering
  \includegraphics[width=0.81\linewidth]{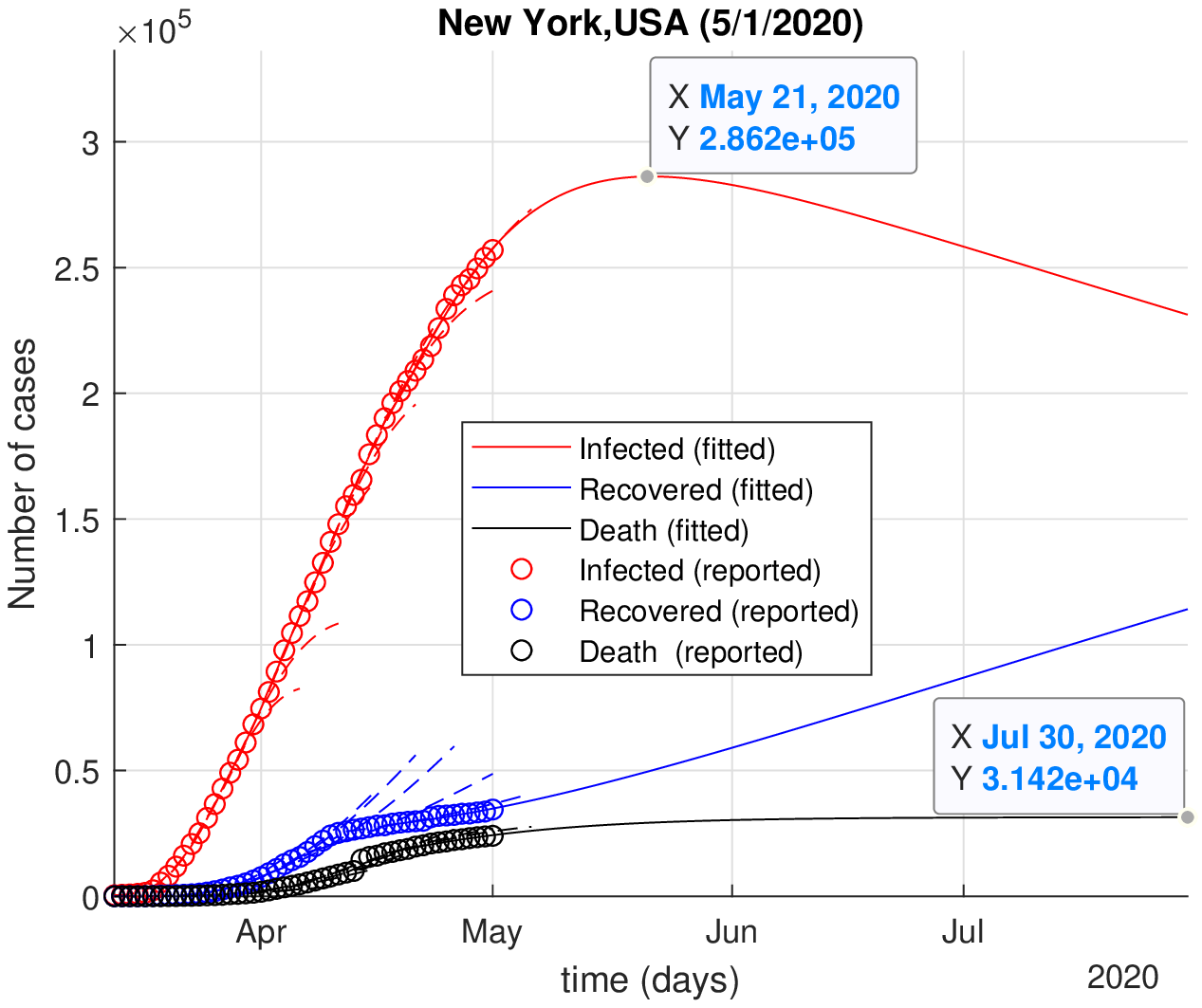}
  \caption{Predictions of the differential SEIRDP model for NewYork state
    from early March to middle July, 2020}
\label{fig:SEIRDP_NewYork_IRD}
\end{figure}

\begin{figure}[!ht]
  \centering
  \includegraphics[width=0.81\linewidth]{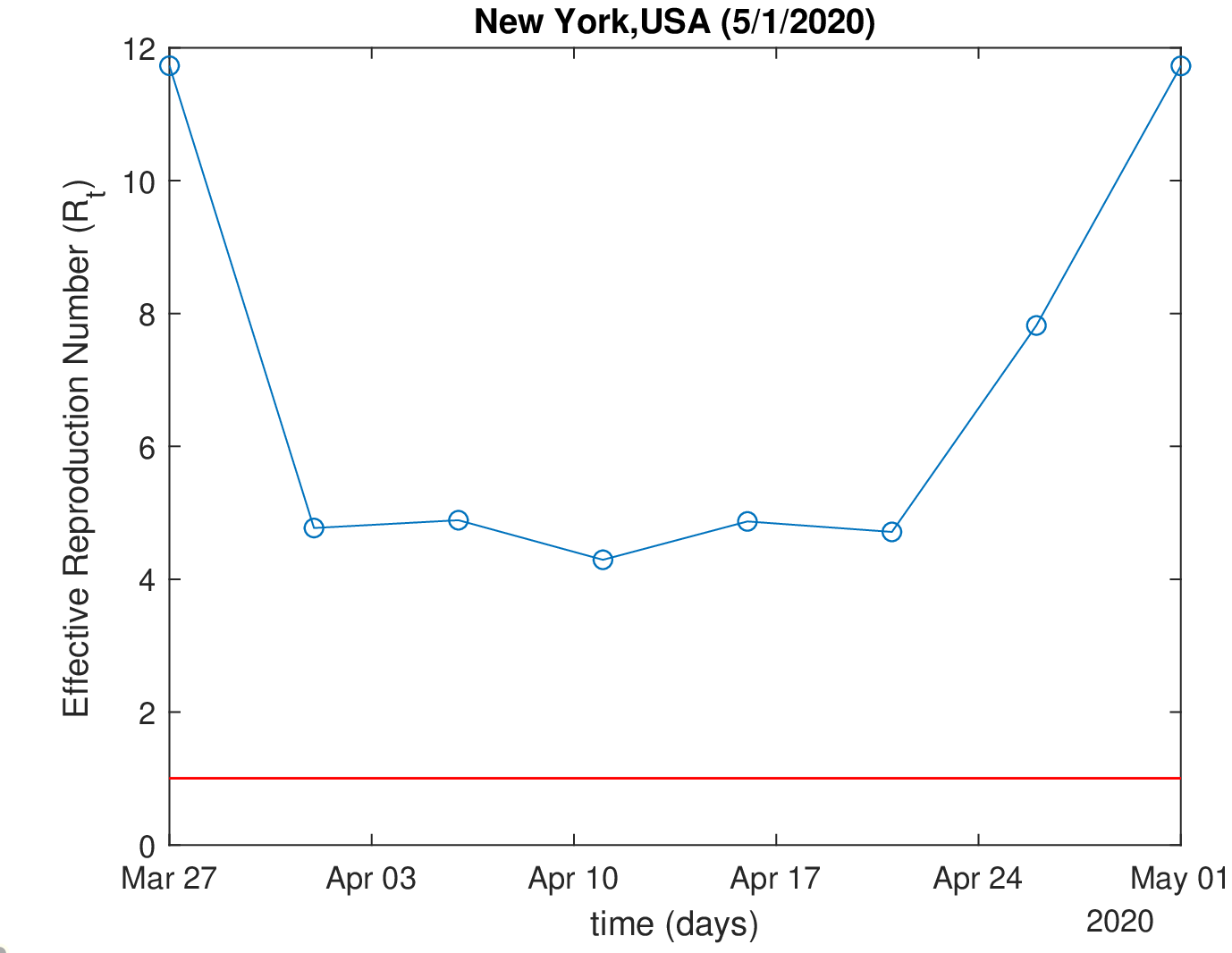}
  \caption{Effective reproduction number for NewYork state from early March
    to middle April, 2020}
\label{fig:SEIRDP_NewYork_BRN}
\end{figure}

\begin{figure}[h]
\centering
\includegraphics[width=0.81\columnwidth]{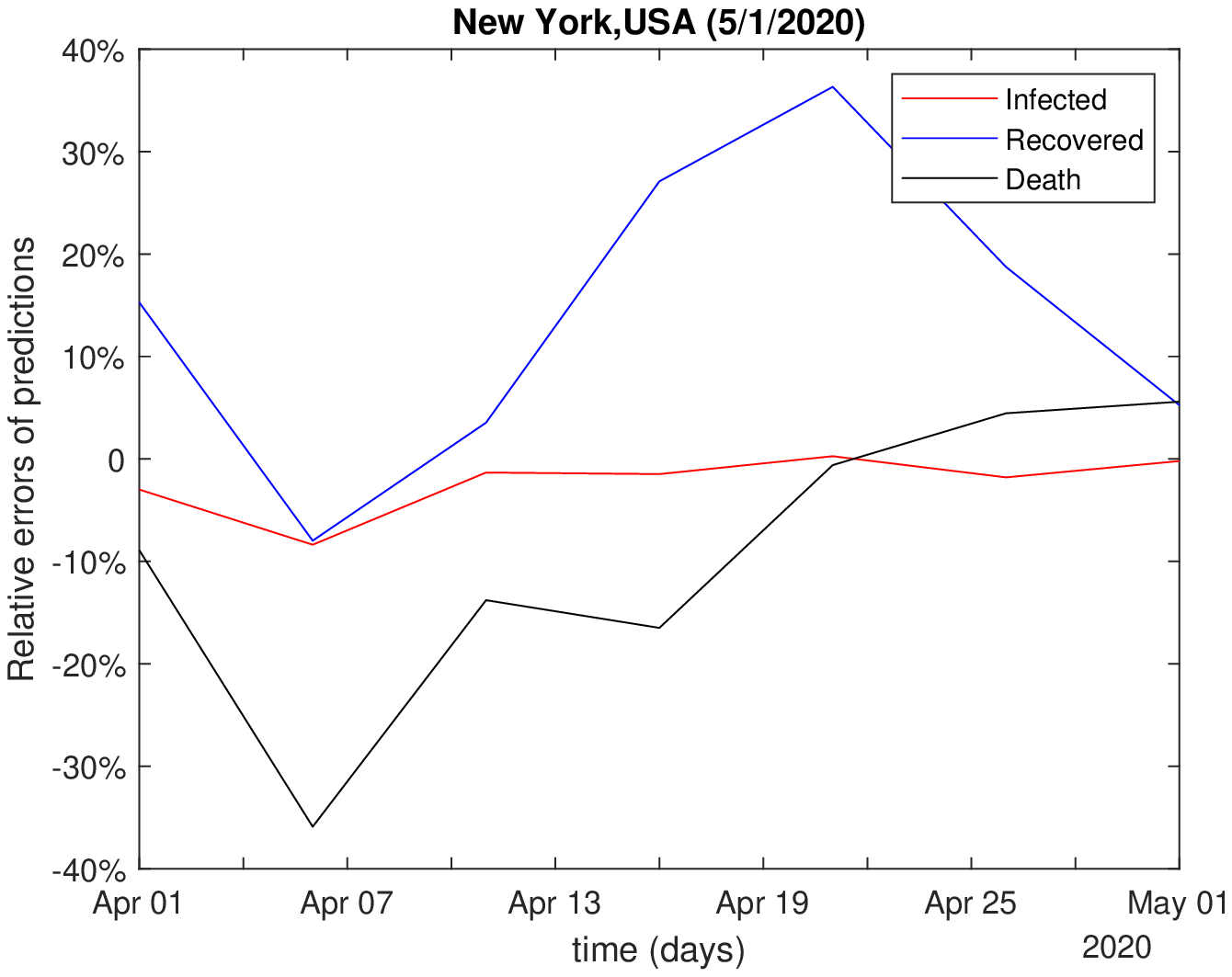}
\caption{The relative mean errors of projection for
  New York state  from early March to middle April,  2020}
\label{fig:SEIRDP_NewYork_error}
\end{figure}

For New York State, the modeling and projection results from the same
period of US are shown in Fig.~\ref{fig:SEIRDP_NewYork_IRD},
Fig.~\ref{fig:SEIRDP_NewYork_BRN} and
Fig.~\ref{fig:SEIRDP_NewYork_error}.
% For New York state, the projected
% currently infected cases  will reach to the peak around the beginning
% of May, which is close to the peak estimation of US. This is no
% surprise, as half of the infected cases come from New York state from
% US. In contrast, IHME's projection shows the time for peak medical
% resource usage is April 8, which indicates significant difference
% from our model.
The estimated effective reproduction numbers from the
differential models are reduced to the be less than 5 and continue
going downward around April 13, 2020.

\begin{figure}[h]
\centering
\includegraphics[width=0.81\columnwidth]{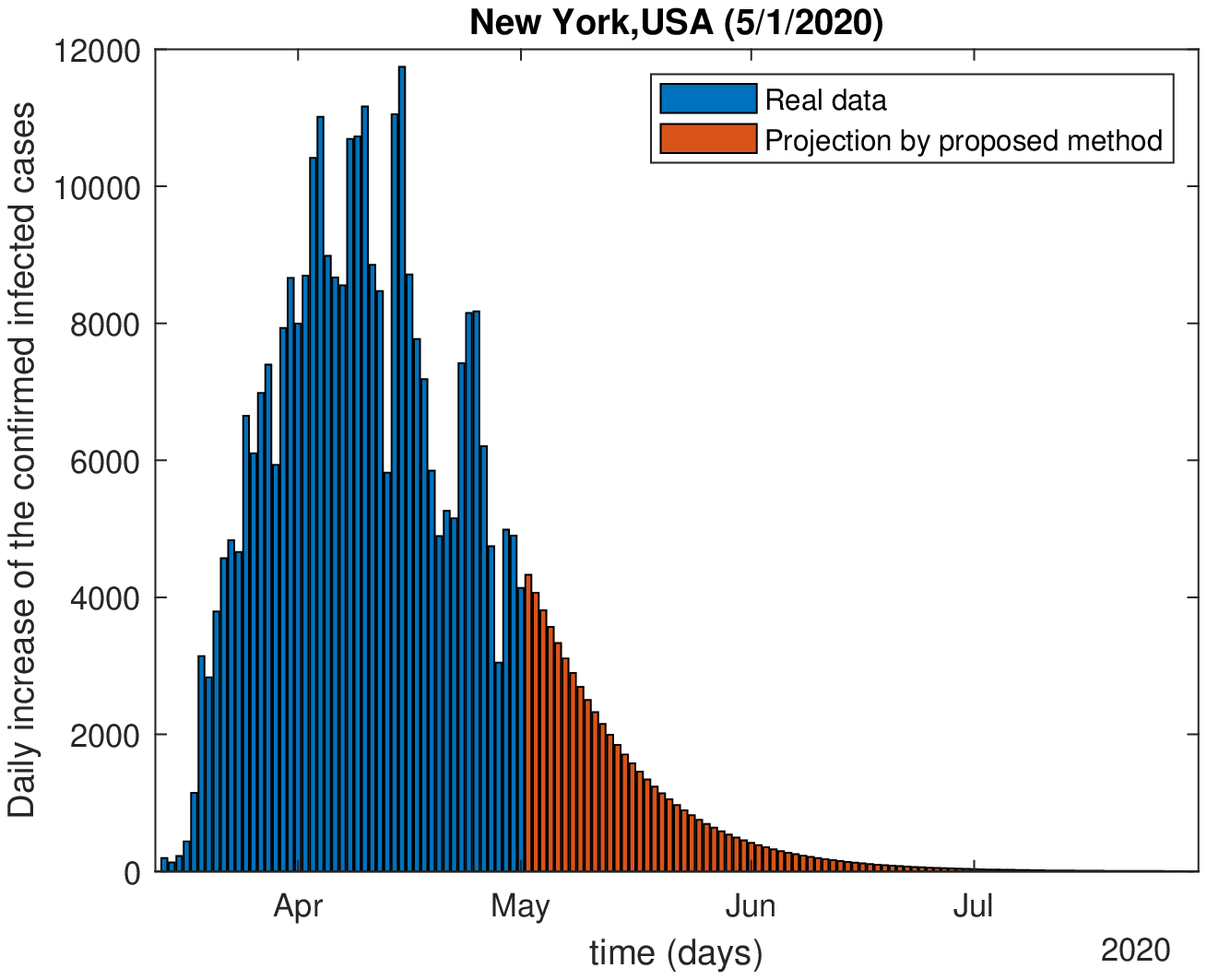}
\caption{The daily increase of infected cases measured and projected
  for NewYork from early Match to middle July, 2020}
\label{fig:SEIRDP_NewYork_NewCase}
\end{figure}

\begin{figure}[htb]
\centering
\includegraphics[width=0.81\columnwidth]{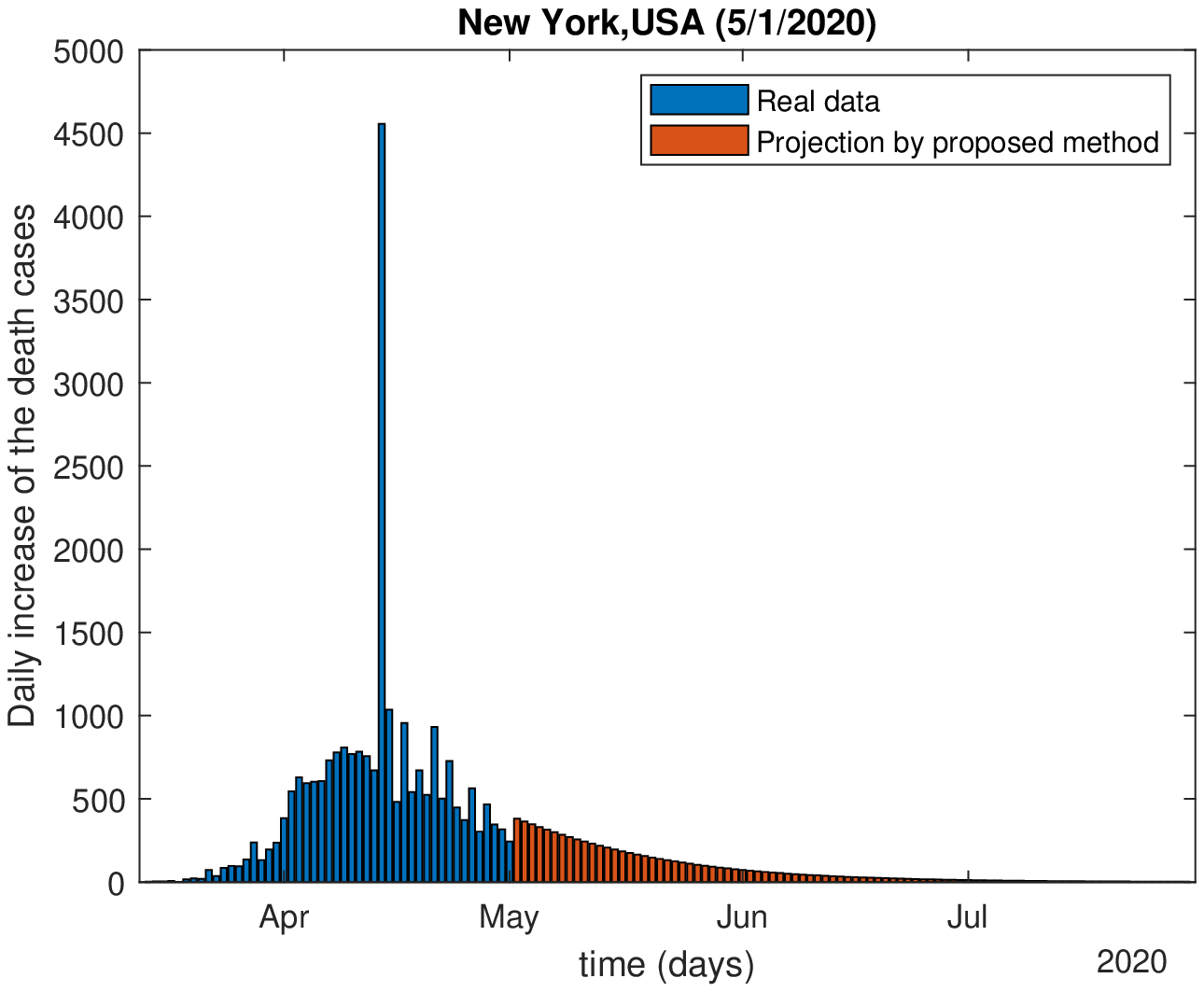}
\caption{The daily increase of death cases measured and  projected for NewYork from early
  Match to middle July, 2020}
\label{fig:SEIRDP_NewYork_NewDeath}
\end{figure}

\begin{figure}[htb]
\centering
\includegraphics[width=0.81\columnwidth]{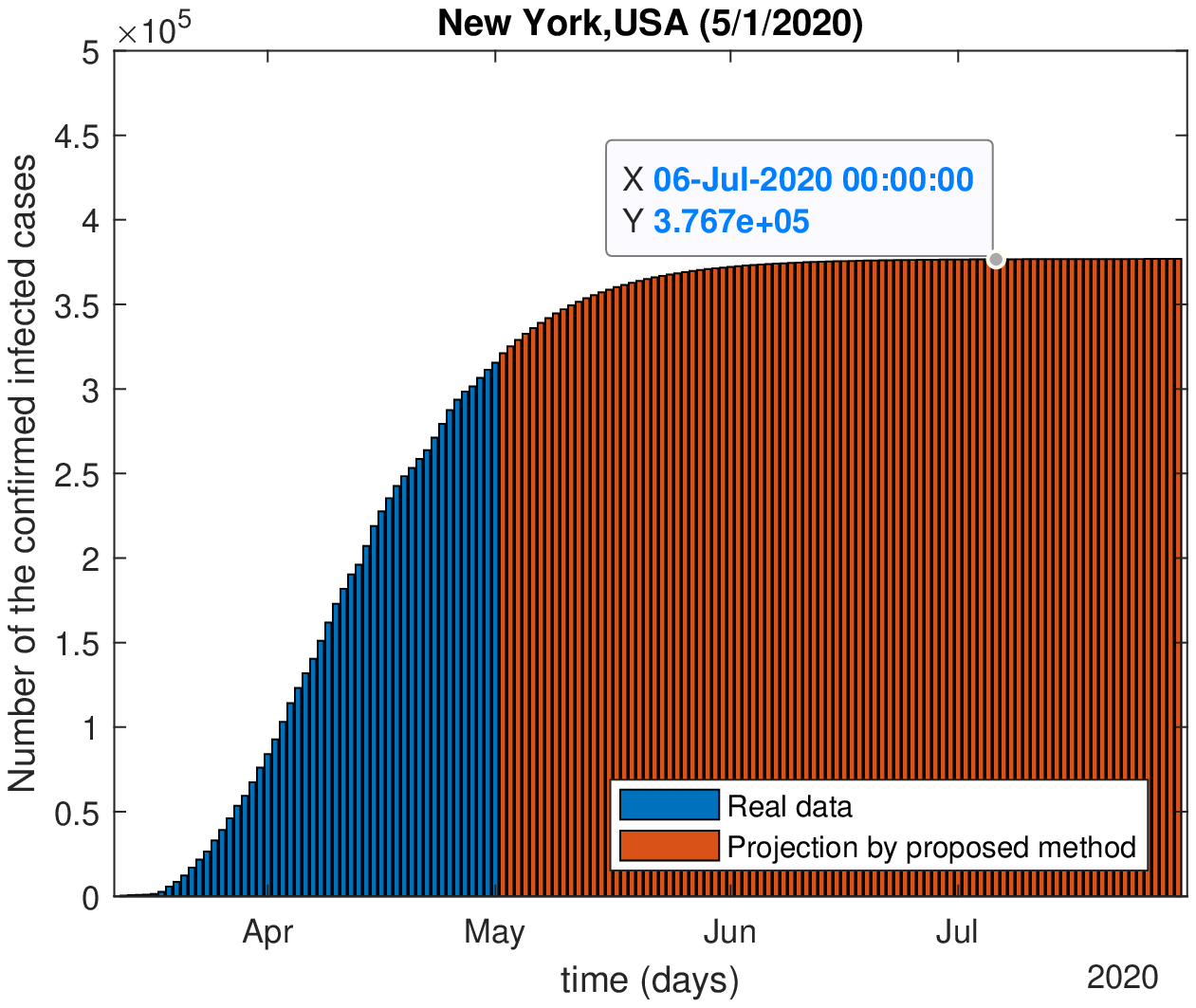}
\caption{The number of the confirmed cases measured and projected for NewYork from early
  Match to middle July, 2020}
\label{fig:SEIRDP_NewYork_Confirmed}
\end{figure}

Fig.~\ref{fig:SEIRDP_NewYork_NewCase},
Fig.~\ref{fig:SEIRDP_NewYork_NewDeath} and
Fig.~\ref{fig:SEIRDP_NewYork_Confirmed} show the daily confirmed
infected cases, the death cases and the total accumulative infected
cases for New York state over mentioned period respectively.
% New York
% state reached the peak around April 10 for both the daily infected
% increase (about 12K per day). But for daily death cases, it will reach
% the peak of 932 per day around April 22.  But the total infected cases
% is projected to reach peak of 311.6K on June 15, and the estimated
% total death cases will reach to about 54.5K in the end.

% \begin{figure}[!ht]
%   \centering
%   \subfigure[]{\includegraphics[width=0.49\linewidth]{SEIRD_NewYork_IRD}\label{fig:SEIRD_NewYork_IRD}}
%   \subfigure[]{\includegraphics[width=0.49\linewidth]{SEIRD_NewYork_BRN}\label{fig:SEIRD_NewYork_BRN}}
%   \caption{(a)Predictions of the SEIRD model and (b)Effective reproduction number for NewYork. 
%   \label{fig:NewYork_SEIRD}}
% \end{figure}

\section{Conclusion}
In this paper, we have proposed a new real-time differential virus
transmission model, which can give more accurate and robust short-term
predictions of COVID-19 transmitted infectious disease with benefits
for near-term trend projection. The new model is based on enhanced
Susceptible-Exposed-Infected-Removed (SEIR) virus transmission
model. As the parameters of the improved SEIR models are trained by
short history window data for accurate trend prediction, our
differential epidemic model, essentially are window-based time-varying
SEIR model. Numerical results on the recent COVID-19 data from China,
Italy and US, California and New York states have been analyzed.

% Based on the projection as of April 13, 2020, the US will reach the
% peak in terms of daily infected cases and death cases around middle
% of April and the actively infected cases will reach the peak around
% the beginning of May, which is also the peak medical resource usage
% time.  The total cumulative infected cases will reach to peak around
% June 12, 2020 with 1.09 million people and the estimated total death
% cases will reach to 90K in the end.

\subsection{Acknowledgment}
The authors would like to thanks Dr. Cheynet for his open-sourced SEIR
model~\cite{cheynet:2020} and for his comments for our work, which
improves the presentation of the article. 

\bibliographystyle{IEEEtranS}

\bibliography{../../bib/epidemic,../../bib/building_model_sim,../../bib/stochastic,../../bib/simulation,../../bib/modeling,../../bib/reduction,../../bib/misc,../../bib/architecture,../../bib/mscad_pub,../../bib/thermal_power,../../bib/neural_network,../../bib/reliability,../../bib/python_libs}

\end{document}